# Editing Cavendish: Maxwell and *The Electrical Researches of Henry Cavendish*

Isobel Falconer

University of St Andrews

## Abstract

During the last five years of his life, 1874-79, James Clerk Maxwell was absorbed in editing the electrical researches of Henry Cavendish, performed 100 years earlier. This endeavour is often assumed to be a work of duty to the Cavendish family, and an unfortunate waste of Maxwell's time. By looking at the history of Cavendish's papers, and the editorial choices that Maxwell made, this paper questions this assumption, considering the importance of Cavendish's experiments in Maxwell's electrical programme, and the implications that he may have derived for developing a doctrine of experimental method.

## Introduction

In 1871 James Clerk Maxwell was elected to the newly established Chair of Experimental Physics at Cambridge, head of the new laboratory that William Cavendish, seventh Duke of Devonshire, had gifted to the University. Three years later, in 1874 Maxwell acquired the unpublished electrical papers of the Duke's relative Henry Cavendish (1731-1810), and undertook to edit them for publication. Over the next five years, he devoted much of the time that he could spare from establishing the laboratory to transcribing and editing the papers. They were published in October 1879, the month before Maxwell died, in 'a classic of scientific editing, locating Cavendish within his own period and – by reporting experimental tests of his results and recasting his ideas into a modern idiom – relating Cavendish's work of the 1770s to the physics of the 1870s'.[1]

By investigating Maxwell's acquisition of the Cavendish papers, and the choices he made in editing them, this paper explores their role in the development of mathematical physics, helps to ascertain why Maxwell devoted time to such a seemingly unimportant task, and questions the assumption that this was purely a work of duty to the Cavendish family.[2]

## Cavendish's work

Henry Cavendish conducted his electrical experiments between 1771 and 1781. During this time he published two papers on electricity in the *Philosophical Transactions of the Royal Society.* The first, published in 1771, was, 'An attempt to explain some of the principal phænomena of electricity by means of an elastic fluid'.[3] Adopting a one-fluid model of electricity, Cavendish was the first person to distinguish clearly between the quantity of electricity in a body (roughly equivalent to charge in modern terms) and the 'degree to which a body is electrified' (akin to potential). He tested the theory against measurements of the charges of bodies of a wide variety of sizes and shapes (i.e. of different capacity) showing that when different bodies were connected together electrically and hence had the same degree of electrification, they carried different charges. 'The ratio of these charges was therefore physically meaningful and, Cavendish showed, measurable.'[4]

---

[1] Cavendish, Henry, *Electrical Researches of Henry Cavendish*, ed. by James Clerk Maxwell (Cambridge University Press, 1879); Maxwell, James Clerk, *The Scientific Letters and Papers of James Clerk Maxwell*, vol.3, ed. by P. M. Harman (Cambridge University Press, 2002), on p12.

[2] This assumption is frequently made, for example, by Harman in *Scientific Letters and Papers,* vol.3, p11, and by A. Whittaker, *James Clerk Maxwell: Perspectives on his Life and Work*, ed. by R. Flood, M. McCartney & A. Whittaker (Oxford University Press, 2013) p116.

In his second paper, of 1776, Cavendish recounted his attempts to imitate the effects of the torpedo (a type of electric fish) by electricity. This paper was a response to considerable debate, initiated by John Walsh in 1773, over whether the shocks delivered by a torpedo were electrical in origin. Those opposed to the idea argued that, if the torpedo was electrified, they did not see, '…why we might not have storms of thunder and lightening in the depths of the ocean.' Guided by his 1771 theory, Cavendish suggested that the shock could be explained by discharge of a very large quantity of electricity, but at a very feeble degree of electrification, and constructed a model torpedo with which he demonstrated this effect to colleagues. He alluded to experiments, that he never published, showing that, 'sea water, or a solution of one part of salt in 30 of water conducts 100 times, and a saturated solution of sea-salt about 720 times better than rain water.'[5]

Maxwell remarks that, 'Such was the reputation of Cavendish for scientific accuracy, that these bare statements seem to have been accepted at once, and soon found their way into the general stock of scientific information.'[6] A similar status was accorded to the 20 packets of unpublished electrical researches that Cavendish left behind on his death in 1810. Information on their contents was scanty to non-existent, but they were believed to contain important results. Although, as Heilbron has shown, even Cavendish's published work was generally neglected, the existence of the papers lingered in the background consciousness of electrical scientists. They were, for example, mentioned by Thomas Young in his 'Life of Cavendish' of 1816.[7] When Maxwell examined them in 1874 he found: a number of drafts for a book on electricity, of which the 1771 paper was to form the first part; and journals recording the details of a large number of experiments and observations as they were made, comparative analysis of the results of the different experiments, and a draft paper. The experiments included a proof of the inverse square law of electrostatic attraction (pre-dating Coulomb's experiment), and long series on the capacitance of different sized and shaped objects, on the effect of coatings on the capacitance of plates (anticipating Faraday's discovery of specific inductive capacity), and on the resistance of salt solutions at different concentrations and temperatures.

---

[3] Cavendish, Henry, 'An Attempt to Explain some of the Principal Phæaenomena of Electricity by Means of an Elastic Fluid,' *Philosophical Transactions of the Royal Society,* 61 (1771) 584-677, repr. in *Electrical Researches,* pp3-63

[4] *Electrical Researches,* p45; Jungnickel, Christa, and Russell McCormmach, *Cavendish* (Philadelphia, Pa: Amer Philosophical Society, 1996) p185.

[5] Cavendish, Henry, 'An account of some attempts to imitate the effects of the torpedo by electricity,' *Philosophical Transactions of the Royal Society,* 66 (1776) 196-225, repr. in *Electrical Researches* pp195-215; Walsh, John, 'Of the Electric Property of the Torpedo. In a Letter from John Walsh, Esq; F. R. S. to Benjamin Franklin, Esq.', *Philosophical Transactions*, 63 (1773), 461–80; Extract from MS. letter of W. Henly, dated 21 May, 1775, in the Canton Papers in the Royal Society's Library, as reported in Maxwell (1879), p.xxxvii; *Electrical Researches,* p195.

[6] *Electrical Researches,* pp.lvi-lvii.

[7] Heilbron, J. L., *Electricity in the 17th and 18th Centuries: A Study of Early Modern Physics* (University of California Press, 1979) p.484; Young, Thomas, 'Life of Cavendish', Supplement to the *Encyclopedia Britannica* 1816-1824, repr. in *The scientific papers of the Honourable Henry Cavendish FRS* ed. by J. Larmor (Cambridge University Press, 1921), 435-448.

## Maxwell's acquisition of the papers

When Cavendish died, childless, in 1810, his papers passed into the hands of his cousin, the fourth Duke of Devonshire, and initially down the family line. However, at some point prior to 1849, the Earl of Burlington, heir to the Devonshire title, put the electrical researches into the hands of William Snow Harris, one of the most prominent British electrical scientists of the day. On Harris' death in 1867 the whereabouts of the papers became obscure. Figure 1 shows the network of connections and influences through which Maxwell obtained them.

*Figure 1. The network of contacts and influences through which Maxwell acquired Henry Cavendish's electrical papers*

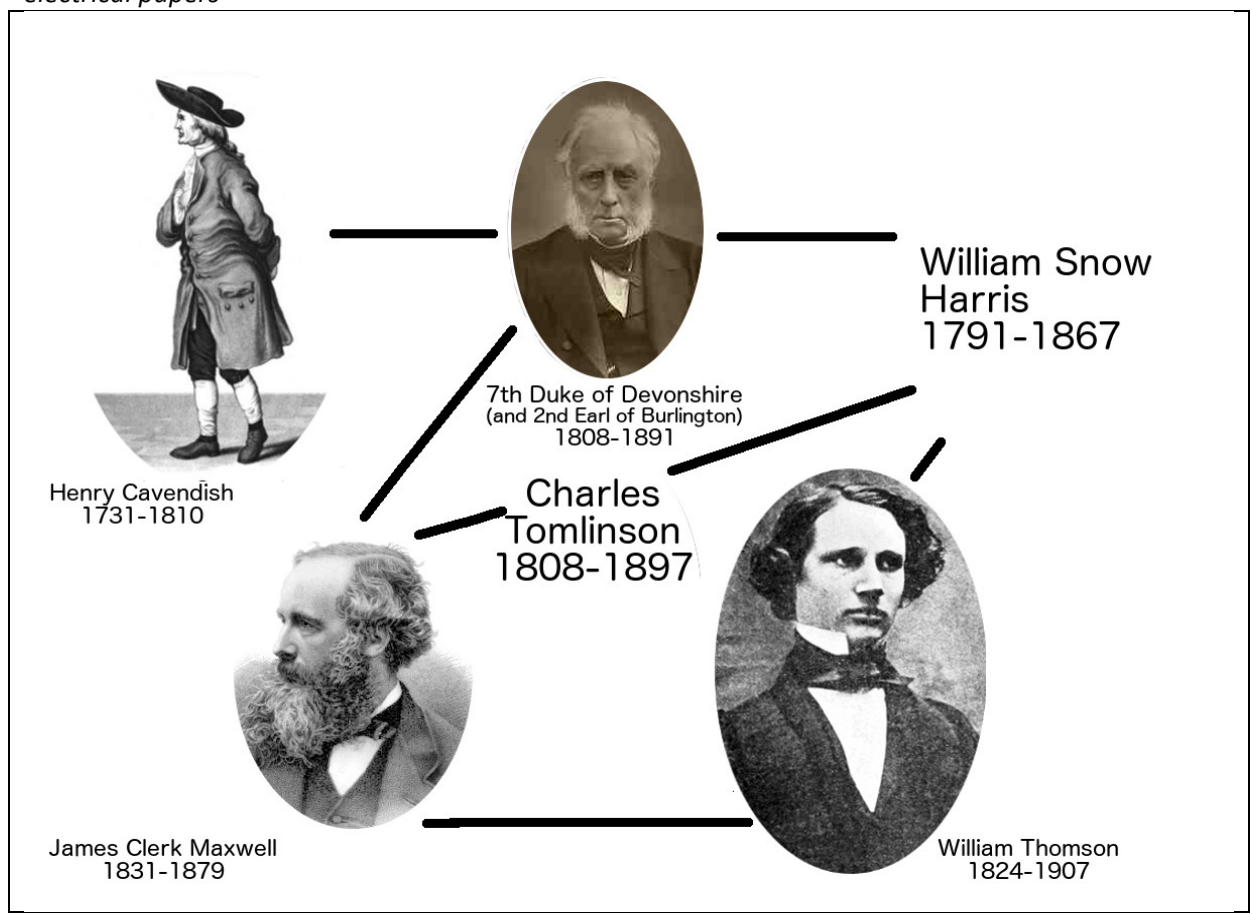

But why did Maxwell want to acquire the papers? To understand this we need to take a step back, to the mid 1830s, and the origins of a simmering debate between Harris and the young William Thomson (later Lord Kelvin).

William Snow Harris is best remembered for his work on lightening conductors, especially on ships, for which he was knighted in 1847. His related programme of experiments on the theory of high tension, static, electricity, was reported between 1834 and his Bakerian Lecture of 1839.[8] In this work, for which he received the Copley medal of the Royal Society in 1835, Harris attempted, '… by operating with large statical forces… to avoid many sources of error inseparable from the employment of very small quantities of electricity, such as those affecting the delicate balance used by Coulomb.'[9] He called into question the prevailing conception of a material electric layer on conductors, due to Poisson, and the generality of Coulomb's inverse square law of electrostatic repulsion, which he found applicable only in situations where induction might change the charge distribution.[10] Instead he suggested that in general, repulsion varied as the direct inverse of distance.

---

[8] James, Frank A. J. L., 'Harris, Sir William Snow (1791–1867)', *Oxford Dictionary of National Biography*, Oxford University Press, 2004; online edn, Jan 2008 [http://www.oxforddnb.com/view/article/12430, accessed 27 April 2015]; Harris, W. Snow, 'On Some Elementary Laws of Electricity', *Philosophical Transactions of the Royal*

Six years later, in 1845, William Thomson, newly graduated from Cambridge and working briefly in Regnault's laboratory in Paris, responded to Liouville's request for clarification of the issues raised by Harris' (and ultimately more importantly, Faraday's) challenges to the laws of electrostatics. His recent reading of Green's little known paper on the uses of potential theory had equipped Thomson with mathematical methods for treating electrostatic theory observationally – using differential equations for macroscopic quantities that could be measured and interpreted using potentials - and avoiding many of the contradictions and problems introduced by the various microscopic hypotheses of electric layers or material electric fluids. However, the applicability of Green's theorem to electrostatics depended upon the correctness of the inverse square law, which Thomson was impelled to defend vigorously. He accordingly criticized Harris' results as due either to disturbing influences or unjustifiable generalization. 'In the experiments made by Mr Harris, we find that no precautions have been taken to avoid the disturbing influence of extraneous conductors, which, according to the descriptions and drawings he gives of his instruments, seem to exist very abundantly in the neighbourhood of the bodies operated upon….'[11]

Despite this critique, Thomson and Harris remained on cordial terms. Beginning in 1847 they corresponded about the relation between spark length and electrostatic force, and in 1849 Thomson visited Harris in Plymouth. While there Harris gave him a brief sight of the Cavendish papers, as he recorded in his notebook. 'Plymouth, Mond., July 2, 1849 Sir William Snow Harris has been showing me Cavendish's unpublished MSS., put in his hands by Lord Burlington, and his work upon them; a most valuable mine of results. I find already the capacity of a disc (circular) was determined experimentally by Cavendish as 1/1.57 that of a sphere of same radius. Now we have capacity of disc … =a/1.571!'[12]

This sight was enough to convince Thomson that the papers contained experimental results that would further his measurement-based electrical programme. Several times over the next 20 years he urged the importance of the papers. In 1851, Cavendish's biographer, George Wilson, recorded Thomson's view that that the papers contained, 'descriptions of excessively ingenious experiments

---

*Society of London* 124 (1834) 213–45; 'Inquiries Concerning the Elementary Laws of Electricity. Second Series', *Philosophical Transactions of the Royal Society of London* 126 (1836) 417–52; 'The Bakerian Lecture: Inquiries Concerning the Elementary Laws of Electricity. Third Series', *Philosophical Transactions of the Royal Society of London* 129 (1839) 215–41, on p215.
[9] Harris, 'Bakerian Lecture', p.215
[10] Buchwald, Jed Z. 1977. 'William Thomson and the Mathematization of Faraday's Electrostatics', *Historical Studies in the Physical Sciences* 8 (1977) 101–136.
[11] Green, George, *An Essay on the Application of Mathematical Analysis to the Theories of Electricity and Magnetism* (Nottingham, 1828), repr. in *The Mathematical Papers of George Green* (New York: Chelsea: 1970), pp. 3-115; Thomson, William, 'On the Mathematical Theory of Electricity in Equilibrium', repr. from the *Cambridge and Dublin Mathematical Journal* Nov. 1845, and from *Phil Mag* 1854 with additional notes dated March 1854, in *Papers on Electrostatics and Magnetism* (London: Macmillan, 1872); Thomson, *Papers on Electrostatics and Magnetism,* p21.
[12] Cambridge University Library, Kelvin collection, Add7342 H36, H37, H38, NB34.

leading to important quantitative results, with reference to electricity in equilibrium on bodies of various forms and dimensions,' and that they should be published. In 1854 Thomson wrote to James Forbes, Professor of Natural Philosophy at Edinburgh, 'I am disposed to regard Cavendish as the founder of the Mathematical Theory of Electricity… I have almost as little doubt but that Cavendish's unpublished papers contain the most accurate measurements that have been made at all on electricity in equilibrium, as that they contain the first accurate measurements that were ever made. Do you think any thing could be done to get them published?' Also in 1854, when revising his 1845 paper on electricity in equilibrium, he raised Cavendish's profile by inclusion of a note on the inverse square law, 'Cavendish demonstrates mathematically that if the law of force be any other than the inverse square of the distance, electricity could not rest in equilibrium on the surface of a conductor.... Cavendish considered the second proposition as highly probable, but had not experimental evidence to support this opinion, in his published work.'[13]

In the meantime, though, private relations between Thomson and Harris were deteriorating. In 1861 Harris drew on and discussed Cavendish's unpublished results on coated plates in his paper, 'On some new phenomena of residuary charge, and the law of exploding distance of electrical accumulation on coated glass'.  Thomson refereed this for Stokes (Secretary to the Royal Society), 'I am considerably bored by a paper of Sir W. S. Harris' which you have sent me. It is so bad, like all he has done – that it would not be creditable to England & the RS, except that.... there are curious & so far as I know novel results of long & varied observation which are worth publishing.' While Harris, whose *Frictional Electricity* utilised more of Cavendish's results and was published posthumously in 1867, wrote in a vitriolic preface clearly aimed at mathematical physicists, '…many profound writers, distinguished for analytical skill, betray an amount of prejudice not very favourable to the advancement of science…. Very little, if any, really useful knowledge of nature is found in the elaborate and interminable pages of symbolic analysis.... As specimens of mere analytical skill they are no doubt valuable, but for any practical result they are frequently valueless.'[14]

Harris died in 1867 and by 1869 Maxwell, who was writing his *Treatise on Electricity and Magnetism*, was making efforts to find out what had happened to Cavendish's papers. Maxwell's correspondence with Thomson during 1868 and 1869 is full of discussion of the capacities and potentials of systems of conducting cylinders, discs and globes, and it seems probable that Thomson alerted him to the relevance of Cavendish's unpublished results. In January 1869, Thomson, in the midst of efforts to establish experiment-based mathematical theory as the proper approach to the development of electrical technology, wrote his 'Determination of the distribution of electricity on a circular segment of plane or spherical conducting surface, under any given influence.' In a footnote he reproduced the memo of his 1849 visit to Snow Harris, along with an adjuration that, 'It is much to be desired that those manuscripts of Cavendish should be published complete; or, at all events, that their safe keeping and accessibility should be secured to the world.'[15] It is possible that Thomson took an even more direct hand, canvassing Maxwell as an intermediary, for Maxwell would have known Charles Tomlinson, to whom he wrote in 1869 enquiring the whereabouts of the papers. Tomlinson was not only Lecturer in Science at King's College School, within the College precinct on the Strand during Maxwell's tenure of the Chair in Natural Philosophy at Kings, but also Harris' friend and collaborator, who prepared his *Frictional Electricity* for publication. Furthermore, he was a Council member of the Cavendish Society, which existed to promote publication of major works in chemistry and had, in 1851, commissioned George Wilson's biography of Cavendish.[16]

---

[13] Wilson, George, *The Life of the Honble Henry Cavendish* (London: Cavendish Society, 1851) p469; Cambridge University Library, Kelvin collection, Add7342 F213; Thomson, *Papers on Electrostatics and Magnetism,* p24.

[14] Harris, W. Snow, 'On Some New Phenomena of Residuary Charge, and the Law of Exploding Distance of Electrical Accumulation on Coated Glass', *Proceedings of the Royal Society of London* 11 (1860) 247–57; Thomson to Stokes 1861, repr. in D. B. Wilson ed. *The Correspondence between Sir George Gabriel Stokes and Sir William Thomson, Baron Kelvin of Largs*, vol.1(Cambridge University Press, 1990) p275; Harris, W. Snow, *A Treatise on Frictional Electricity: In Theory and Practice* (London: Virtue, 1867) pxxiii.

[15] Maxwell, James Clerk, *The Scientific Letters and Papers of James Clerk Maxwell,* vol.2, P. M. Harman, ed. (Cambridge University Press, 1995); Smith, Crosbie, and M. Norton Wise, *Energy and Empire: A Biographical Study of Lord Kelvin* (Cambridge University Press, 1989); Thomson, W., 'Determination of the distribution of electricity on a circular segment of plane or spherical conducting surface, under any given influence' (dated January 1869) repr. in *Papers on Electrostatics and Magnetism*. p180

It is worth emphasizing that Maxwell initiated enquiries for the papers two years before he had any personal connection with the Cavendish family, and that the evidence suggests that his interest originated directly from his electromagnetic programme and correspondence with Thomson.

It took four years, but in 1873, Maxwell was able to write triumphantly to Thomson, 'The Tomlinson Correspondence is found.' Apparently Harris' son had the papers, and had resisted suggestions that they be put in the hands of the Royal Society. By this time, however, Maxwell knew, and had consulted with, the Duke of Devonshire over plans for the new laboratory in Cambridge. Now he enlisted the Duke's help. 'In the interest of science and at the suggestion of several scientific men I write to ask your help in securing the preservation of those manuscripts of Henry Cavendish which relate to electricity…. [and which] were put into the hands of Sir William by the Earl of Burlington…. Many men of science are naturally anxious that the preservation of papers so important should not depend on the accidents attendant on the transmission of such manuscripts from hand to hand and all such anxiety would be removed if your Grace whom I understand to be the representative both of the Hon Henry Cavendish and of the Earl of Burlington were to take steps to obtain the papers from Mr Harris.'[17] Maxwell's innocence of the aristocracy is betrayed here by his evident ignorance that the current Duke of Devonshire and the Earl of Burlington were one and the same person.

Although in March 1873 Maxwell reported to Thomson that, 'The Chancellor is now fairly engaged to collect the Cavendish papers,' the younger Harris was apparently reluctant to give them up. Once again Maxwell appealed to Tomlinson, '… as the person most likely to be able to render assistance.' At last the Duke received the papers and, by July 1874, had placed them in Maxwell's hands, presumably with a view to publication.[18]

Maxwell reported every stage of the recovery of the papers to Thomson, to whom he also confided that, 'I am just going to walk the plank with them for the sake of physical science.' The duty expressed here is to physical science, rather than to the Cavendish family as benefactors of the Cambridge laboratory. A review in *Nature* in 1873 attributed to Maxwell makes clear the possible value of Cavendish's results in his and Thomson's electrical programme, '… in the last century Henry Cavendish led the way in the science of electrical measurement, and Coulomb invented experimental methods of great precision…. Then came Poisson and the mathematicians, who raised the science of electricity to a height of analytical splendour.... And now that electrical knowledge has acquired a commercial value, and must be supplied to the telegraphic world in whatever form it can be obtained, we are perhaps in some danger of forgetting the debt we owe to those mathematicians who… [represented] qualities which we now know to be capable of direct measurement, and which we are beginning to be able to explain to persons not trained in high mathematics.' In a comparable passage in the preface to his *Treatise on Electricity and Magnetism* Maxwell points out that, 'The important applications of electromagnetism to telegraphy have also reacted on pure science by giving a commercial value to accurate electrical measurements.'[19]

---

[16] Kurzer, F. 2004. 'The Life and Work of Charles Tomlinson FRS: A Career in Victorian Science and Technology'. *Notes and Records of the Royal Society,* 58 (2) 203–26.
[17] *Scientific Letters and Papers* vol.2, p784; p785; p785.
[18] *Scientific Letters and Papers* vol.2, p839; p858; *Scientific Letters and Papers* vol.3, p82.

## Maxwell's editing of Cavendish's *Electrical Researches*

Between 1874 and 1879 Maxwell, with the help of William Garnett, the Demonstrator at the Cavendish Laboratory, sorted, transcribed, and prepared the papers for publication. Maxwell rapidly became an enthusiast, declaring Cavendish's methodical account, '… the best piece of scientific writing on the evidence of the exactness of the theory of electricity which has yet been published,' that his methods, '… are unique of their kind even if the date were the corresponding years of this century instead of 1771-2-3,' and that, 'If these experiments had been published in the authors life time the science of electrical measurement would have been developed much earlier.'[20] He sought out old instruments, delved into eighteenth century chemical nomenclature, tested Cavendish's method of judging conductivity, repeated and improved his inverse square law experiment, compared many of Cavendish's results with more recent ones and drew on them in refereeing papers, and utilised some of the results in his own papers and the second edition of his *Treatise*.[21]

However, like all editors, Maxwell made decisions about what to include, what to leave out, how to represent it, and what was worthy of comment. By exploring some of these decisions we gain an appreciation of what he was trying to achieve in editing the papers.

### What to include

Maxwell found that, 'the mathematical part and the description of the experiments is in a much more finished state than I had thought,' and that Cavendish himself had prepared much of it for publication – why he had not published remains a mystery. These, there was no question, should now be published, along with reprints of the two papers from the *Philosophical Transactions*, because, '… everyone has not the Philosophical Transactions of that year.' Equally unproblematic seems the decision to exclude earlier drafts for the 'more perfect papers.'[22]

The daily journal of experiments was a more difficult problem. 'I do not think that Cavendish would have himself published these and therefore it becomes a question whether it is right to do so now.' Eventually Maxwell decided to publish the journals for 1771 and 1772 in full. His reasons are illuminating. As well as being, '… a decided advantage to the reader… to be able to refer to the details of each experiment,' Maxwell gave methodology, the value of the results, and Cavendish's priority claims as reasons: **'**I do not think any mere statement of the results… would supersede the actual record of the work as an example of method'; '… they contain all the data of some of the most important electrical experiments,' and; '… when we are publishing for the first time his electrical discoveries made a century ago the whole of the evidence becomes of greater importance than it was then.'[23]  The hint here that one of Maxwell's aims was to establish Cavendish's priority is made much clearer in another draft, which also casts light on Maxwell's own approach to publication. 'When an experimentalist publishes his own researches his object is to establish the truth of his discoveries. He therefore explains his experimental methods and states his results but unless the experiments are very difficult and not likely to be repeated he leaves it to others to verify the results by repeating the experiments. But when we are printing for the first time experimental discoveries made a century ago it is not so much the truth of the discoveries that we wish to establish as the fact that Cavendish made these discoveries a century ago, and therefore it becomes desirable to exhibit the whole evidence for this fact.'[24]

---

[19] *Scientific Letters and Papers* vol.2, p839; 'Review of Fleeming Jenkin, *Electricity and Magnetism*', *Nature,* 8 (1873) 42-43, attribution to Maxwell in *Scientific Letters and Papers* vol.2 p842; Maxwell, James Clerk, *Treatise on Electricity and Magnetism,* 1st edn (Oxford: Clarendon, 1873) px.
[20] *Scientific Letters and Papers* vol.3 p373; p383; vol.2 p539.
[21] *Scientific Letters and Papers* vol.3 p531; p718; *Electrical Researches* p417; *Scientific Letters and Papers* vol.3 p472; e.g. J. Clerk Maxwell, 'On the Electrical Capacity of a Long Narrow Cylinder, and of a Disc of Sensible Thickness', *Proceedings of the London Mathematical Society,*  9 (1877-8) 94-101 and *Electrical Researches* p393-400.
[22] *Scientific Letters and Papers* vol.3 pp373-374.
[23] *Scientific Letters and Papers* vol.3 p374; p374; p376; *Electrical Researches* pxliv; *Scientific Letters and Papers* vol.3 p374.

## How to represent the work

As noted above, Harman judged Maxwell's as, 'a classic of scientific editing,' the principles of which include that, '… the reproduction of the texts faithfully follows the manuscript….'[25] Nowhere is this more evident than in Maxwell's concern over reproduction of Cavendish's drawings and diagrams. 'Mr Garnett has made facsimiles of the drawings of the experimental arrangements and Macmillan tells me it would be easy to have these engraved exactly as Cavendish drew them…. In them there must be no conjectural emendations. The geometrical diagrams, however, may be made as clear as we can without attempting to copy any irregularity in Cavendish's pen.'[26]

However, there may have been more to this concern than scholarly correctness. Maxwell's reference to 'no conjectural emendations' opposed his edition directly to William Snow Harris' accounts of Cavendish's work, woven into the argument of Harris' *Frictional Electricity*. This opposition is evident in Figure 2, which compares Harris' diagram of Cavendish' diverging electrometer with Maxwell's 'warts and all' reproduction of Cavendish's sketch of the apparatus.

*Figure 2. Cavendish's diverging pith ball electrometer as represented by Harris (left) and Maxwell (right).[27]*

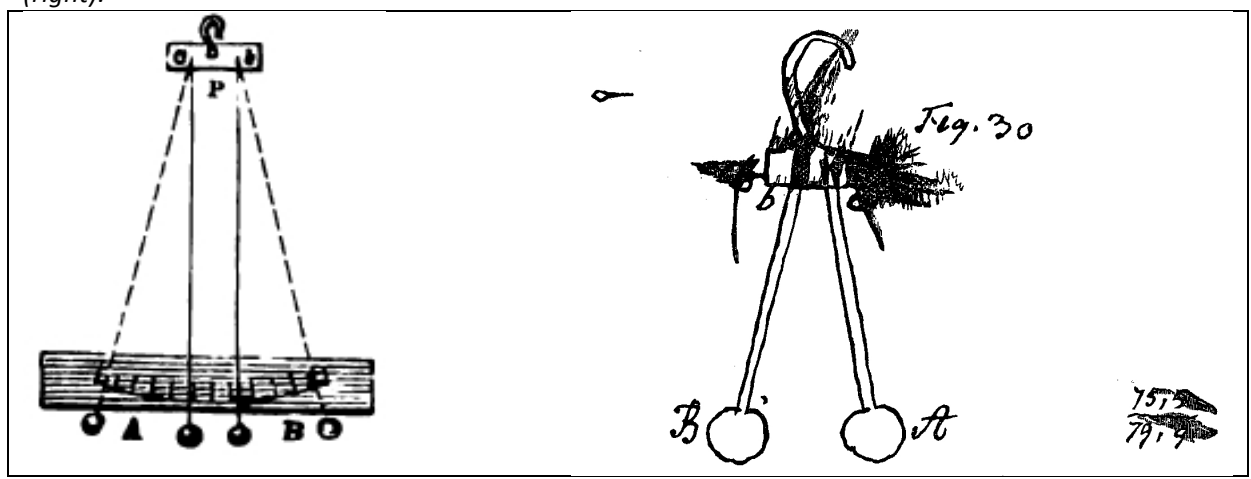

Nowhere was the contrast between Harris and Maxwell more obvious than in their accounts of Cavendish's demonstration that there was no charge inside a hollow spherical conductor, shown in Figure 3 and described in the next section.

*Figure 3. Cavendish's apparatus to show there is no charge inside a hollow spherical conductor, as represented by Harris (left) and Maxwell (centre and right).[28]*

---

[24] Cambridge University Library, Maxwell Collection, Add7655 Vc33.
[25] Harman in *Scientific Letters and Papers,* vol.3 p12; pxxiii.
[26] *Scientific Letters and Papers,* vol.3 pp374-375.
[27] Harris *Treatise on Frictional Electricity* p24; *Electrical Researches* p121.

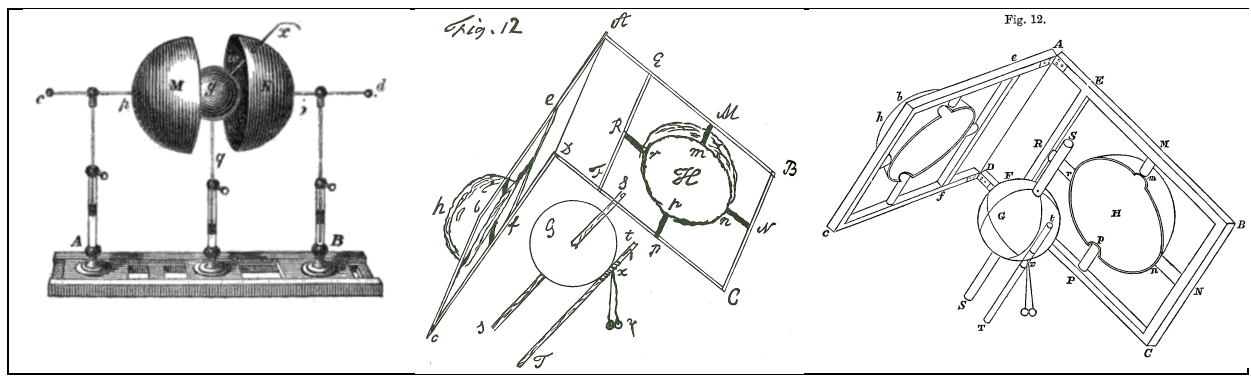

This experiment is an indirect confirmation of the inverse square law and both Cavendish and Maxwell considered it fundamental - so important that Maxwell reproduced it both with the 'irregularities of Cavendish's pen' and as a clear line drawing without them. Harris represented it very differently, with clear 'conjectural emendation,' even though his written description corresponded more nearly to Cavendish's and Maxwell's pictures.

In his diagrams, Maxwell was implicitly asserting the authority of his version of Cavendish over Harris' and, by association, the authority of his approach to electrical science over that of Harris and his like.

## What to comment on
Maxwell included editorial notes on various aspects of Cavendish's work at the end of the book. Five of the topics he chose to comment on are discussed here.

### *Electrical theory*
Cavendish had arrived at his concept of 'degree of electrification' by considering electricity as an elastic fluid that yet, when in a wire connecting two conductors, behaved as though incompressible – a disjunction that he considered the weakest point of his theory. For Maxwell this weakness was insignificant compared to the insight that, as George Green had pointed out, 'The meaning which [Cavendish] here fixes to [the terms positively and negatively electrified], … is equivalent to the meaning of the modern term potential, as used by practical electricians. The idea of potential as used by mathematicians is expressed by Cavendish in his theory of canals of incompressible fluid.'[29]

Although Maxwell was not explicit, the equivalence between 'degree of electrification' and 'potential' was an instrumental one – both were measured in the same way with an electrometer. Without this instrumental equivalence, Maxwell's assertion of mathematical equivalence, which introduces a potential term at the outset in the equilibrium conditions for a conductor and then demonstrates consistency with many of Cavendish's results, and his discard of some of the theoretical points of difference between potential and canals of incompressible fluid, hold little conviction.

Maxwell's cavalier attitude indicates how essential the equivalence was to the whole enterprise. Without it, Cavendish's work would not have served as a precedent and a model for Maxwell and Thomson's electrical programme. Maxwell reinforced the relevance by extended comparisons of Cavendish's experimental measurements of the capacity of various-shaped objects with mathematical calculations by Maxwell and Thomson based on potential theory and Thomson's method of electrical images.

---

[28] Harris *Treatise on Frictional Electricity* p45; *Electrical Researches* p104, 106.
[29] *Electrical Researches* p382

### *No charge inside a hollow spherical conductor*

The inverse square law is a necessary condition for potential theory to be of any use in electrostatics. Prior to the development of potential theory, though, Cavendish, as a convinced Newtonian with an essentially material concept of electric fluid, had other reasons to test for such a law. He demonstrated theoretically that only if the inverse square law is exactly true will the charge on a spherical conductor reside in equilibrium on the surface, with no charge inside the conductor. He published this result in his 1771 paper but, until Harris and Maxwell examined his papers, no one realised that he had also demonstrated it experimentally.[30] As Maxwell pointed out, Cavendish's experiment pre-dated Coulomb's by at least 10 years.[31]

Cavendish placed an insulated conducting globe inside two hollow conducting hemispheres (also insulated), and held in a hinged frame which could be opened to remove the hemispheres from around the globe (see Figure 3). The frame was closed, and a fine wire inserted to connect the globe with the outer sphere. The whole apparatus was electrified, then the connecting wire removed using an attached silk thread, the frame opened, and the outer hemispheres discharged to earth. A pith ball electrometer was used to test the charge on the globe, which was found to be nil. Cavendish calculated theoretically what the charge on the globe would be if the power in the law of repulsion were $-(2+n)$ and estimated that if $n$ were greater than $\pm 1/50$ he would have detected it.[32]

Now Maxwell decided to repeat Cavendish's experiment. His student Donald McAlister did the work using an improved apparatus of Maxwell's design. They encased the inner globe in the outer sphere, except for a small hole through which a wire connecting the globe to the sphere or to the electrometer passed, the hole being covered by a removable cap. This shielded the inside of the apparatus from possible disturbances, and also prevented leakage of charge from the globe, whose insulating supports now rested on the inside of the sphere. They also benefited from the far greater precision of Thomson's quadrant electrometer in estimating that $n$ could be no greater than $\pm 1/21600$.

Maxwell's motives in repeating Cavendish's experiment are obscure. On the face of it, he had no need to do so. Even before he knew that Cavendish, or anyone else, had performed it with any degree of accuracy, he asserted with confidence in 1873 in his *Treatise*, that the generally observed absence of charge on one conductor enclosed within another was a better argument for the inverse square law than were Coulomb's experiments.[33] 'The results, however, which we derive from such experiments [as Coulomb's] must be regarded as affected by an error depending on the probable error of each experiment, and unless the skill of the operator be very great, the probable error of an experiment with the torsion-balance is considerable. As an argument that the attraction is really, and not merely as a rough approximation, inversely as the square of the distance, Experiment VII [showing the absence of charge inside a conductor] is far more conclusive than any measurements of electrical forces can be.'[34]

---

[30] Harman's suggestion that Maxwell's letter to Thomson of 15 October 1864 may refer to the electrostatic experiment is clearly mistaken, since examination of Thomson's notebook and correspondence shows that he did not spot this experiment during his very brief visit to Harris, and that Maxwell would not have known in 1864 that Cavendish had performed it. The phrase 'the Cavendish experiment' was (and is) usually reserved for Cavendish's gravitation experiment, and the drawing Maxwell includes accords more nearly with that experiment. See *Scientific Letters and Papers* vol.2 p179.

[31] *Electrical Researches* pxxxii, xlviii.

[32] *Electrical Researches* p112.

[33] This argument has a long history dating back to Joseph Priestley. See Heilbron *Electricity in the 17th and 18th Centuries.*

[34] Maxwell *Treatise* 1st edn p75.

That Maxwell did develop an experiment that he avowed already so well established, emphasises once again its importance to the whole electrical enterprise. Further, he may have been responding to the electrical standards programme in which he and Thomson were heavily engaged, and in which the inverse square law was implicated, which was putting great emphasis on the precision of standards measurement. Although in 1873 he records no evidence that anyone had tried the experiment other than very crudely, he remarks that, 'The methods of detecting the electrification of a body are so delicate that a millionth part of the original electrification of B [the inner conductor] could be observed if it existed. No experiments involving the direct measurement of forces can be brought to such a degree of accuracy.'[35] Hence Cavendish's method was more precise and less error prone than Coulomb's torsion balance – which, as we have seen, was open to criticisms of the type Harris levelled at it. 'Cavendish thus established the law of electrical repulsion by an experiment in which the thing to be observed was the absence of charge on an insulated conductor. No actual measurement of force was required. No better method of testing the accuracy of the received law of force has ever been devised.' Maxwell's development achieved greater precision still. [36]

### *Electrical properties of non-conductors*
Maxwell devoted a long note to Cavendish's discovery that the capacity of various solid non-conductors was greater than that of air – what Faraday was later to call 'specific inductive capacity'.[37] He called attention to the similarity between Cavendish's conceptual model for such dielectrics, of conducting strata, and his own model proposed in the *Treatise* in 1873 to account for electric absorption.[38] He concluded the note with a comparison of Cavendish's measurements of the specific inductive capacity of various densities of flint glass with several more recent determinations.

What Maxwell does not mention in his note is the context within which he was concerned with dielectrics and proposed his strata model. This can be gleaned by only reading the *Treatise.* In an extended reworking of his section on the 'Physical interpretation of Green's function' in the second edition of the *Treatise*, published in 1881, Maxwell drives home the necessity for invoking his 'displacement current'. 'We have hitherto confined ourselves to that theory of electricity which … takes no account of the nature of the dielectric medium between the conductors. ... But this is true only in the standard medium, which we may take to be air. In other media the relation is different, as was proved experimentally, though not published, by Cavendish, and afterwards rediscovered independently by Faraday. In order to express the phenomenon completely, we find it necessary to consider two vector quantities, the relation between which is different in different media. One of these is the electromotive intensity, the other is the electric displacement.'[39]

This was the point at which Maxwell departed from Thomson, who could never accept the idea of a displacement current and based his own electrical theory on a simple analogy with heat flow. In the *Treatise* Maxwell was explicit that, 'The object of [the strata model] is merely to point out the true mathematical character of the so-called electric absorption, and to shew how fundamentally it differs from the phenomena of heat which seem at first sight analogous.' The two theories differed in their experimental implications, with Maxwell suggesting that values measured for specific inductive capacity depended on the length of time for which the substance was electrified.[40] The point of comparing Cavendish's measurements with those of Hopkinson, Wüllner, Gordon and Schiller, was to argue that while the first three had measuring procedures taking a second or two and given results that were too high, Gordon and Schiller had measured at a rate of 1200 or 14000 interruptions per second respectively, and given results that were quantitatively more in line with Maxwell's theories.  Maxwell did not point out that Hopkinson had performed his experiments under Thomson's aegis, while Gordon had done his under Maxwell's.

---

[35] Smith and Wise *Energy and Empire;* Maxwell *Treatise* 1st edn p74.
[36] *Scientific Letters and Papers* vol.3 p539; for an account of Coulomb's experiment and some of its critics see Falconer, I., 'Charles Augustin Coulomb and the Fundamental Law of Electrostatics', *Metrologia*, 41 (2004) S107-S114; for a critique of the theory of Cavendish's method and its successors see Fulcher, L. P., and M. A. Telljohann. 'On the Interpretation of Indirect Tests of Coulomb's Law: Maxwell's Derivation Revisited'. *American Journal of Physics* 44 (1976) 366–69. doi:10.1119/1.10196.
[37] ER note 15 p402-404
[38] *Electrical Researches* p402-404; Maxwell *Treatise* 1st edn p381.
[39] Maxwell, James Clerk, *Treatise on Electricity and Magnetism*  2nd edn (Oxford: Clarendon Press, 1881) TEM2 p133.

### *Measuring conductivity, and physiology*

Cavendish lived before the invention of galvanometers. Maxwell's highest pitch of enthusiasm was aroused by the methods Cavendish used instead in his experiments on conductivity. 'All these results and many more were got by comparison of the strength of shocks taken through Cavendish's body. I think this series of experiments is the most wonderful of them all, and well worth verification.' He proceeded with gusto to verify the method, conscripting students and visitors alike to try it. Arthur Schuster recalled, 'a young American astronomer expressing in severe terms his disappointment that, after travelling on purpose to Cambridge to make Maxwell's acquaintance and to get some hints on astronomical subjects, the latter would only talk about Cavendish, and almost compelled him to take his coat off, plunge his hands into basins of water and submit himself to the sensation of a series of electrical shocks.'[41]

Cavendish's method, wonderful though it might be, posed a problem for Maxwell. In Cavendish's time self-report of sensation in the experimenter's own body were a commonplace part of a natural philosopher's practice. The credibility of the evidence depended crucially on the credibility of the person reporting their sensations. As Schaffer puts it, 'True philosophers knew themselves. They could be trusted to tell what had happened to them.' Yet by the 1870s individual sensation had become deeply suspect as scientists, including Maxwell himself, attempted to shift the burden of evidence from their own bodies to self-registering instruments.[42] Fleeming Jenkin made this shift very clear in the Introduction to his *Electricity and Magnetism,* published in 1873. He contrasted the science of electricity as portrayed in textbooks with that known to 'practical electricians' such as Maxwell and Thomson. The former contained an, 'apparently incoherent series of facts,' while the latter were more scientific. Jenkin was promoting the latter. Yet, 'Many of the assertions [of practical electricians] cannot be proved to be true except by complex apparatus, and the action of this complex apparatus cannot be explained until the general theory has been mastered.' In a review of the book, attributed to Maxwell, Jenkin's distinction became one between a science of 'sparks and shocks which are seen and felt,' and a science of 'currents and resistances to be measured and calculated.'[43] How then, was Maxwell to enlist Cavendish's results, obtained by shocks, in support of his own electrical measurement programme?

Maxwell pursued a dual approach to this problem. First he investigated whether bodily methods could actually measure quantitatively any parameters such as current that were meaningful in his electrical theory. Cavendish had employed shocks both for qualitative exploration, assessing the strength of sensation when the conditions varied, and for quantitative results when he equated two sets of experimental conditions where the shocks felt equal (see Figure 4). Maxwell tried to ascertain whether shocks could be compared reliably and consistently, and what factors affected the comparison. Perhaps he did not get the answers he expected for, a year later in November 1878, he wrote to the physiologist Ernst Fleischl, 'Perhaps you may be able to tell me if any experiments have been made on the relation between the circumstances of an electric discharge namely its quantity the mean strength of the current and the total quantity which passes, and (1) the effect on a muscle (2) the sensation felt by a man.'[44]

---

[40] Wise, M. Norton, 'The Flow Analogy to Electricity and Magnetism, Part I: William Thomson's Reformulation of Action at a Distance', *Archive for History of Exact Sciences*, 25 (1981) 19–70, on p36; Maxwell *Treatise*  2nd edn p419; *Electrical Researches* p403.
[41] *Scientific Letters and Papers*  vol.3 p530; Schuster, Arthur (1910) in *A History of the Cavendish Laboratory* (London: Longmans Green, 1910) p33
[42] Schaffer, Simon, 'Self Evidence', *Critical Inquiry*, 18 (1992), 327–62, on p329; Morus, Iwan Rhys, 'What Happened to Scientific Sensation?', *European Romantic Review*, 22 (2011), 389–403.
[43] Jenkin, Henry Charles Fleeming, *Electricity and Magnetism*, Text-Books of Science (London: Longmans, Green, and Co., 1873) pv-vi; 'Review of Fleeming Jenkin' p42.

*Figure 4. Cavendish's records of qualitative results of the shock given by his artificial torpedo (left) and quantitative comparison of the conductivity of salt solutions when the shocks felt equal (right). The right-hand figure shows Cavendish's own notebook as reproduced by Maxwell (above) and the printed version of the same entry (below)*[45]

On hearing from Fleischl that, 'The effect of an electric discharge through a nerve does not depend neither on the mean strength nor on the total quantity which passes, but on the rate of change of intensity of the electric current,' Maxwell embarked on his second approach. He turned the experiments on their head, as he had previously done with colour vision, so that instead of using physiological effects to measure physical phenomena, he used physical effects to measure physiology. In March 1879 he planned two experiments, 'on the physiological effect of an induction current,' and 'on the physiological effects of electric discharges.'[46] The results were reported in the *Electrical Researches* note 31. Pointing out that Cavendish had used only transient currents, from the discharge of Leyden jars, Maxwell compared the effects of transient currents with different decay constants. For the induction experiments, 'got by varying the strength of the primary circuit and the resistance in the secondary, and I find that if the resistance of the secondary circuit (including the victim) is as the square of the strength of the primary current, the shock of breaking seems about as intense,' with similar results for the discharge experiments (Figure 5). However, the sensation did vary with the rate of decay. With a very rapid decay but an initial current, '… large enough to produce a shock of easily remembered intensity in the wrists and elbow, there is very little skin sensation,' whereas with a slower decay, '… but still far too small for the duration of discharge to be directly perceived, the skin sensation becomes much more intense… so that it becomes almost impossible so to concentrate attention on the sensation of the internal nerves'The condition of the hands also had implications for how Cavendish's experiments were to be interpreted: 'As the hands get well soaked and seasoned to shocks the pricking goes off more than the nerve shock, so that the index becomes less than 2. Cavendish made it greater than 2 so perhaps his hands were not so wet, and he went more by the 'pricking of his thumbs' than I did.'[47]

---

[44] *Scientific Letters and Papers* vol.3, p716.

[45] *Electrical Researches* p310, p327 and facing.

[46] *Scientific Letters and Papers* vol.3 p717; Harman, P. M., *The Natural Philosophy of James Clerk Maxwell* (Cambridge University Press, 2001); *Scientific Letters and Papers* vol.3 p759-763.

*Figure 5. Maxwell's plan for an experiment on the physiological effects of an electric discharge. The 'victim' in the centre assessed alternately the effects of discharge of condenser $K_1$ charged to potential $V_1$ through resistance $R_1$, and condenser $K_2$ charged to potential $V_2$ through resistance $R_2$. The initial strength of the current is V/R, and the time modulus of decay is KR*[48]

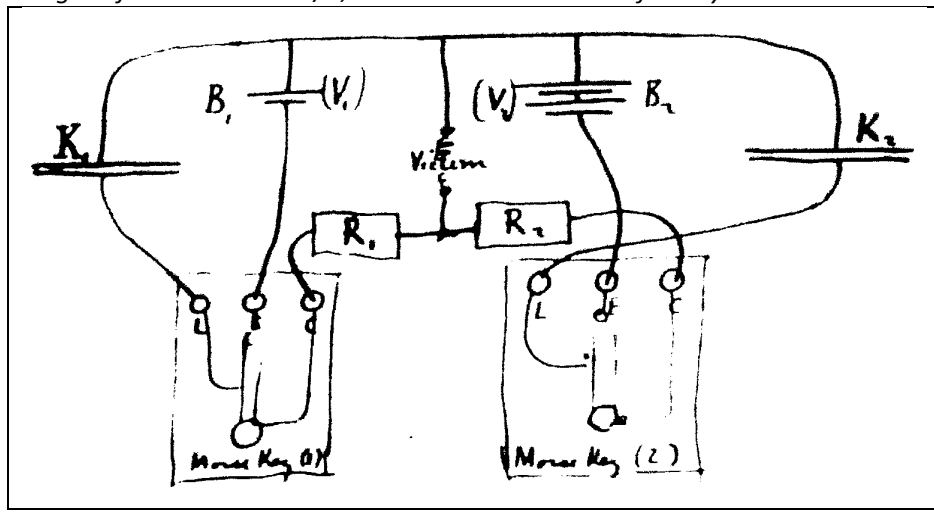

In these physiological experiments the objectification of the body is made very evident by Maxwell's continued use of the term 'victim' to describe the person sensing the shocks, which occurs in his letters, his published account, and his diagrams (see Figure 5). Moreover, he left the results here, as measures of the body's response to electric shocks. He did not go back and re-evaluate the implications for Cavendish's measurements of conductivity – although this is possibly due to his rapidly deteriorating health and the fact that he had already sent the bulk of the book for printing. Had he lived longer he might have taken the topic, or its implications, further.

Both these approaches might be considered as pursuing Maxwell's ambition to develop the 'doctrine of method,' outlined as the proper work of a physics laboratory in his inaugural lecture at Cambridge.[49]

---

[47] *Scientific Letters and Papers* vol.3 p764; *Electrical Researches* p439; *Scientific Letters and Papers* vol.3 p764
[48] Cambridge University Library, Maxwell Collection, Add 7655 Vc32.
[49] Maxwell, James Clerk, 'Introductory Lecture on Experimental Physics' (inaugural lecture as professor of experimental physics at Cambridge), in W. D. Niven ed. *The Scientific Papers of James Clerk Maxwell* vol.2 (Mineola NY: Dover, 2003) p250

### Ohm's Law

In one of the most problematic passages in the *Electrical Researches* Maxwell claimed that Cavendish had discovered Ohm's Law well before Ohm. This is one of his earliest observations about the papers, writing to Garnett in July 1874 that Cavendish, '… made a most extensive series of experiments on the conductivity of saline solutions… and it seems as if more marks were wanted for him if he cut out G.S.Ohm long before constant currents were invented.' He repeated the claim in 1877, 'Cavendish is the first verifier of Ohm's Law, for he finds by successive series of experiments that the resistance is as the following power of the velocity, 1.08, 1.03, .980, and concludes that it is as the first power,' and with similar wording in his introduction to the book.[50]

At first glance this assertion is surprising, since we are more used to the form *potential = resistance x current* and, assuming Cavendish used a constant discharge potential, and that 'velocity' was somehow comparable to current, we might expect resistance to be inversely rather than directly as the power of the velocity.

We need to examine carefully what was important to Maxwell about Ohm's law, as well as what Cavendish actually did. Since 1863 Maxwell had been heavily involved in the British Association Committee on Electrical Standards, and in establishing an absolute standard for resistance. Ohm's law was essential here for establishing the relationship between potential and current.[51] In 1873, in the *Treatise,* Maxwell wrote that, 'The introduction of this term [resistance] would have been of no scientific value unless Ohm had shewn, as he did experimentally, that it corresponds to a real physical quantity, that is, that it has a definite value which is altered only when the nature of the conductor is altered,' and hence the resistance does not vary when the magnitude of the current varies. Yet in 1874, Schuster questioned this result and the British Association set up a committee to investigate. Chrystal and Saunders' investigation of resistance at a wide range of current intensities, directed by Maxwell, was the first major experimental project in the Cavendish Laboratory. Thus, in commenting on Cavendish, Maxwell was keen to stress that Cavendish had found resistance to be independent of current. 'The resistance … varies as the 0.08, 0.03, -0.024 power of the strength of the current in the first three sets of experiments, and in the fourth set that it does not vary at all.'[52]

To arrive at this statement Maxwell explained that by 'resistance' Cavendish meant, 'the whole force which resists the current, and by "velocity" the strength of the current through unit of area of the section of the conductor,'[53] whereas in his, Maxwell's parlance, 'resistance' meant the force which resists a current of unit strength. This explanation enabled him to reduce the power of the velocity in Cavendish's results by one, arriving at the conclusion that resistance was independent of the current.

However, Maxwell's statement that, 'By four different series of experiments on the same solution in wide and in narrow tubes, Cavendish found that the resistance varied as the 1.08, 1.03, 0.976, and 1.00 power of the velocity'[54] is misleading for two reasons. First, Cavendish performed only a single experiment in 1773, with a further one in 1781. Second, slips in Cavendish's calculation, which Maxwell apparently overlooked, invalidate the equivalence of the results.

---

[50] *Electrical Researches* p334; *Scientific Letters and Papers* vol.3 p82; p530; *Electrical Researches* plix.

[51] Smith and Wise pp 687-690; Schaffer, Simon, 'Late Victorian Metrology and Its Instrumentation: A Manufactory of Ohms,' in Bud and Cozzens ed. *Invisible Connections: Instruments, Institutions, and Science* (Bellingham: SPIE, 1992).

[52] Maxwell *Treatise* 1st edn; Harman in *Scientific Letters and Papers* vol.3 p10; *Electrical Researches* plix-lx.

[53] *Electrical Researches* plix.

[54] *Electrical Researches* plx.

In all his comparisons of conductivity or resistance, Cavendish took the length of the tube of salt solution as a measure of resistance. He calibrated his tubes by filling them with mercury, measuring the length of the column and then pouring the mercury out and measuring its weight. He took the weight per inch of mercury (proportional to the cross-sectional area) as his measure of velocity, and thus was assuming that the time taken for each discharge in the comparison was the same. On Cavendish's model that electricity moved through a conductor as an incompressible fluid, 'velocity' was thus proportional to the quantity of electricity moving through the solution in the time of the discharge.

In November 1773 Cavendish performed a single measurement of 'what length of a tube, 37 inches of which held 44 grains [of water], the shock must pass, so as to be as much diminished as in passing through 44¼ of the large one.' Cavendish estimated that the large, wider, tube held 250 grains in 37 inches. He judged that the two shocks were equal when the discharge passed through 6.8 inches of the narrower tube. Thus, he said,

$$\frac{6.8}{44¼} = \frac{44}{250}\bigg]^{1.08}$$

and concluded that, 'the resistance should seem as the 1.08 power of the velocity.'[55]

He subsequently re-calibrated the tubes more accurately using mercury, and re-calculated his result to give the resistance as the 1.03 power of the velocity. But he had not repeated the experiment.

Cavendish returned to this question seven years later, in January 1781. He compared two different tubes (tube 15, which held 7.7 grains of mercury in 11.55 inches, and tube 5 which held 489 grains in 42.1 inches).[56]

He judged that the shock through 2.75 inches of the narrow tube (15) felt the same as through 41.9 inches of the wide tube (5). Using the same two tubes he repeated the measurement and this time judged 2.85 inches of tube 15 equivalent to 41.9 of tube 5. This was a repeat reading under the same experimental conditions that could have been averaged to give a mean reading.  However, Cavendish listed the two results separately as giving resistance as the 0.976 power of velocity, and resistance 'directly as velocity,' and Maxwell took them at face value as a series of results, adding a triumphant footnote, 'This is the first experimental proof of what is now known as Ohm's law.'[57]

Although in both cases Cavendish concluded that the resistance was approximately directly as the velocity, with a power close to one, examination of his actual working reveals that in 1773 was the inverse of that in 1781. In concluding as he did in 1773 from the equation above, Cavendish is using the length of the tube as a measure of resistance, and the weight of fluid in 37 inches as a measure of velocity, as outlined above. His equation amounts to,

$$\frac{resistance_1}{resistance_2} = \left(\frac{velocity_1}{velocity_2}\right)^n$$

Yet in 1781, his calculation is of,

$$\frac{length_1}{length_2} = \left(\frac{velocity_2}{velocity_1}\right)^n \text{ i.e. } \frac{resistance_1}{resistance_2} = \left(\frac{velocity_2}{velocity_1}\right)^n$$

In other words, in 1773 his measurements gave resistance as approximately proportional to velocity, whereas in 1781 they gave resistance as approximately inversely proportional to velocity.

---

[55] *Electrical Researches* p294; p294.
[56] *Electrical Researches* p337.
[57] *Electrical Researches* p333-334.

In his 1781 journal Cavendish subsequently defined resistance as inversely proportional to the weight in grains per inch of mercury in the tube, which accords with his 1781 calculation.[58]

Perhaps Maxwell was getting carried away by enthusiasm. Unlike the physiology experiments, where his untimely death might explain why he did not re-examine Cavendish's results, the Ohm's law results were ones he had been commenting on for five years. The persistence with which Maxwell overlooked the problems with the 'series' of experiments and calculations is a strong indication of his commitment to Cavendish's priority in this discovery.

## Conclusion

As historians we can use Maxwell's editing of Cavendish's *Electrical Researches* as a lens to examine his personal scientific situation and his perception of the state of electrical science in the 1870s. He had just become the first Professor of Experimental Physics at Cambridge and in this role he was committed to 'forming a school of scientific criticism, and in assisting the development of the doctrine of method.'[59] His work on Cavendish, and in particular his improvements to the null method of the inverse square law experiment, and his lengthy investigation of the reliability of Cavendish's bodily methods, might be seen as part of this endeavour.

But beyond the bounds of Cambridge, it was clear that Maxwell and Thomson did not consider the battle for the hearts and minds of electrical scientists won with the success of the transatlantic cable and the publication of Maxwell's *Treatise on Electricity and Magnetism.* Maxwell's editing of Cavendish can be read as a move in this battle*.* Thomson had realised the potential value of the papers as early as 1849, and the pair went to some effort to acquire them. In his editorial decisions, Maxwell took deliberate, though never explicit, aim at electrical scientists like Snow Harris and his followers, who did not have a proper appreciation of mathematical electrical theory and hence did not understand the proper precautions or measurements to be taken during experiments. At the outset of the enterprise, Maxwell and Thomson were clearly working together. But this did not prevent Maxwell, on occasion, from highlighting those of Cavendish's ideas and results that might promote his own views of electromagnetism over those of Thomson. Again, such opposition was implicit; Maxwell's was a partisan account that did not mention the existence of an alternative view.

By emphasising Cavendish's skill as an experimentalist, while claiming continuity between their theories, Maxwell provided an experimental genealogy for his own electrical programme – one that might appeal to 'practical men' without much mathematics. This genealogy was, above all, British, exemplified by his priority claims for Cavendish in the discovery of both Coulomb's law and Ohm's law.

However one reads it, Maxwell's *Electrical Researches of the Honourable Henry Cavendish* was more than just a labour of duty to the Cavendish family.

---

[58] *Electrical Researches* p337.
[59] Maxwell 'Introductory Lecture' p250.